\documentclass[12pt]{article}
\usepackage{epsfig}
\def\vbs{\vspace{2cm}}

\font\grande=cmr10 scaled \magstep4
\font\medio=cmr10 scaled \magstep2

\def\laq{\raise 0.4 ex \hbox{$<$}\kern -0.8 em\lower 0.62 ex\hbox{$\sim$}}
\def\gaq{\raise 0.4 ex \hbox{$>$}\kern -0.7 em\lower 0.62 ex\hbox{$\sim$}}

\def\laq{\raise 0.4ex\hbox{$<$}\kern -0.8em\lower 0.62
ex\hbox{$\sim$}}
\def\gaq{\raise 0.4ex\hbox{$>$}\kern -0.7em\lower 0.62
ex\hbox{$\sim$}}

\begin{document}
\bibliographystyle {unsrt}
\newcommand{\pa}{\partial}

\titlepage
\begin{flushright}
DF/IST--2.98 \\
January  1999\\
\end{flushright}
\vspace{15mm}

\begin{center}
{\bf Cosmological Solutions of Type II Superstrings}\\
\vspace{5mm}
{\grande }
\vspace{10mm}
M. C. Bento\footnote{Also at Centro Multidisciplinar de Astrof\' \i sica (CENTRA), Instituto Superior T\'ecnico, Lisbon, Portugal. }

{\em Departamento de F\'{\i}sica,
  Instituto Superior T\'ecnico}\\ 

{\em Av. Rovisco Pais, 1096 Lisboa Codex, Portugal } \\
\end{center}

\vspace{10mm}
\centerline{\medio  Abstract}

\vspace{0.5cm}
\noindent
We study maximally symmetric cosmological solutions of type II supersymmetric strings in the presence of the exact quartic curvature corrections to the lowest order effective action, including loop and D-instanton effects. We find  that, unlike the case of type IIA theories,  de Sitter solutions exist for type IIB superstrings,  a conclusion that remains valid  when  higher-curvature corrections are included   on the basis of SL(2,Z) invariance.  
\\

\vfill

\newpage

\setcounter{equation}{0}
\setcounter{page}{2}

Recently, there has been considerable interest on the subject of cosmological solutions of the low-energy limit of string theories and M-theory.
String theories, in their low-energy limit, give rise to effective theories of gravity containing higher-derivative corrections.
In type II superstring theories there are no corrections quadratic or cubic in the curvatures so that the first corrections beyond the Einstein-Hilbert term are fourth order in the Riemann curvature, arising at order $\alpha'^3$. These  terms  have been computed explicitely via string four-point amplitude calculations \cite{gross} and confirmed  in $\sigma$-model perturbation theory, from a four-loop divergence that contributes to the $\beta$-functions and gives $\alpha'^3$ corrections to the effective action \cite{grisaru}. For four gravitons, in particular, there exists a one-loop result for the quartic-curvature corrections \cite{sakai} and  all contributions higher than one loop  vanish  due to non-renormalization theorems  \cite{tseytlin}.

  There exist also non-perturbative corrections to the $R^4$ term due to  D-instantons. Superstring  dualities, relating superstring perturbative expansions (and M-theory), provide strong constraints on the non-perturbative structure of string theories \cite{bachas}. In type IIB theory, in particular, non-perturbative effects are intimately related to the SL(2,Z) symmetry of the theory \cite{hull} and a conjecture for their structure has been put forward in Ref.  \cite{green},  which gives the  tree and one-loop corrections to the $R^4$ term besides the instanton ones; this conjecture has been validated in \cite{antoniadis}.
In type IIA theories such non-perturbative corrections do not exist, essentially because there are no finite-action instantons so that the $R^4$ term is given entirely by the sum of the perturbative tree-level and one-loop terms. 

 The effect of higher-curvature terms in the string  low energy effective actions on the maximally symmetric cosmological solutions of the theory has been studied, with the result that, for all string theories, at  tree-level, there are  de Sitter solutions but they are unstable and these solutions disappear once the dilaton is included \cite{bento1}. 
The aim of this letter is to study the effect  of loop   and non-perturbative corrections at all orders in $\alpha'$, in type II superstrings, on this type of solutions. 

We start by considering the bosonic effective Lagrangian of the theory at lowest order in $\alpha'$ (hereafter we set $\alpha'=1)$ in the Einstein frame 

\begin{equation}
\label{aa}
{\cal L}_0=R - {1\over 2 \phi_2^2} \partial_\mu S \partial^\mu {\bar S} - {1\over 12 \phi_2} (S H^1 + H^2)_{\mu\nu\rho} ({\bar S} H^1 + H^2 )^{\mu\nu\rho},
\end{equation}
where $S=\phi_1 + i~ \phi_2= \chi + i e^{-\phi}$ and  $H_{\mu\nu\rho}^{1,2}$ is the antisymmetric tensor field strength (in the following we shall set $<H^{1,2}>=0)$. This Lagrangian is invariant under an  SL(2,R) symmetry that acts as

\begin{equation}
\label{ab}
g_{\mu\nu}\rightarrow g_{\mu\nu}, \quad S \rightarrow {aS + b \over c S + d}, \quad B_{\mu\nu}^a\rightarrow (\Lambda^T)^{-1 a}_b B^b_{\mu\nu}, \quad \Lambda=\left(\begin{array}{cc} a & b\\ c & d  \end{array}\right)\in SL(2,R).
\end{equation}

A discrete subgroup of this symmetry, $SL(2,Z)$, is conjectured to be an exact non-perturbative symmetry of the type IIB string \cite{schwarz}.

The first perturbative  corrections to the effective Lagrangian \cite{gross} arise at order  $\alpha'^3$ and are given by

\begin{eqnarray}
\label{aca}
{\cal L}_3&= &{\zeta(3) \over 3. 2^6} e^{\frac{3}{2}\phi} ({t_8}^{\mu_1 \mu_2 \ldots\mu_8} {t_8}^{\nu_1 \nu_2 \ldots\nu_8} + {1\over 8} {\epsilon_{10}}^{\mu_1\mu_2 \ldots\mu_8\mu_9\mu_{10}}  {{\epsilon_{10}}^{\nu_1 \nu_2 \cdots \nu_8}}_{\mu_9\mu_{10}})\times \nonumber\\
 & &{\hat R}_{\mu_1\mu_2\nu_1\nu_2}{\hat R}_{\mu_3\mu_4\nu_3\nu_4}{\hat R}_{\mu_5\mu_6\nu_5\nu_6} {\hat R}_{\mu_7\mu_8\nu_7\nu_8} +\ldots\nonumber\\
&\equiv & {\zeta(3) \over 3. 2^6} e^{\frac{3}{2}\phi}(t_8 t_8 + \frac{1}{8} \epsilon_{10}\epsilon_{10}) R^4,
\end{eqnarray}
where $\zeta$ is the Riemann $\zeta$-function,

\begin{equation}
\label{ad}
{{\hat  R}_{\mu\nu}}^{\alpha\beta}= {R_{\mu\nu}}^{\alpha\beta} - {1\over 4}g_{[\mu}^{[\alpha} \nabla_{\nu]}\nabla^{\beta]}\phi ,
\end{equation}
 $t_8$ is the standard eight-index tensor arising in string amplitudes,  $\epsilon_{10} $ is the totally antisymmetric symbol in ten dimensions and $\ldots$ denotes terms involving derivatives of the dilaton. This Lagrangian reproduces the four-point amplitude calculated in string theory and it is in agreement with  $\sigma$-model calculations.

Firstly, we shall look for  maximally symmetric solutions of  ${\cal L}_0 +  {\cal L}_3 $ (provisionally we set $\chi=0$).  For  maximally symmetric spaces

\begin{equation}
\label{af}
 R^\mu_{\phantom{\mu} \nu\lambda\sigma}=K\left(
\delta^\mu_\lambda g_{\nu\sigma} - \delta^\mu\sigma g_{\nu\lambda}\right), \qquad \phi=\phi_c=cte.
\end{equation}

In this case, the  metric and dilaton equations  of motion reduce to the following algebraic equations

\begin{eqnarray}
\label{afe}
\alpha K + \beta K^4 = 0,\\
\beta K^4 =0,
\label{afg}
\end{eqnarray}
 where $\alpha=(D-1)(D-2)$ and $\beta=- {3\over 8}\zeta(3) e^{-3/2 \phi} (D-3)(D-2)(D-1) $. We consider first the case $\phi=0$, for which only eq.~(\ref{afe}) is relevant. Clearly, Minkowski space is a solution and, with our conventions, de Sitter (anti-de Sitter) solutions correspond to real positive (negative) roots of (\ref{afe}). In the critical number of dimensions, D=10, there is only one positive  root, $K_s=\left(- { \alpha \over \beta} \right)^{1/3}$.  However, this solution is unstable since stability of solutions depends on the sign of $\delta^2 {\cal L}$ and this is  determined by the sign of $f'(K_s)$ \cite{deser}, where $f(K)=\alpha K + \beta K^4$, which is negative.    If $\phi\not= 0$, eq.~(\ref{afe}) has to be taken into account as well and it is clear that, in this case, de Sitter space is no longer a solution and only Minkowsky space remains.
  
At the perturbative level, there exist string one-loop corrections to the four-point functions. For four gravitons, these corrections have been calculated and amount to the exchange

\begin{equation}
\label{ah}
\zeta(3)\rightarrow \zeta(3)+{\pi^2 \over 3} \phi_2^{-2}
\end{equation}
in eq.~(\ref{aca}). The solution of the metric equation of motion  now becomes $K_s=\left(- {\alpha \over {\hat \beta}}\right)^{1/3}$, where ${\hat \beta }= -{3\over 8} e^{-{3\over 2}\phi}(\zeta(3)+{\pi^2\over 3} e^{-2\phi}) (D-3)(D-2)(D-1)$, again a de Sitter solution. On the other hand, the dilaton equation of motion changes to ${\hat\beta}K^4=0$, which again has only Minkowsky space as a solution.

 Non-renormalization theorems ensure that there are no higher-loop corrections.
On the other hand, in type IIB theory there can be non-perturbative contributions due to  D-instantons \cite{green}. The multi-instanton contributions are determined by the SL(2, Z) symmetry, i.e. the exact non-perturbative result for four gravitons can be found  by covariantizing the perturbative result, which is not SL(2, Z)-invariant \cite{green}

\begin{equation}
\label{ai}
{\cal L}_{3}={1\over 3.2^7} f_0(S, \bar S) \left(t_8 t_8 + {1\over 8} \epsilon_{10} \epsilon_{10} \right) R^4,
\end{equation}
where we have used the compact notation defined in  eq.~(\ref{aca}); $f_0(S,{\bar S})$ is the non-holomorphic modular form  such that $f_0(S, {\bar S})=\zeta(3)E_{3/2}(S)$, where $E_s(S)$  is a non-holomorphic Eisenstein series defined by

\begin{equation}
\label{aja}
E_s(S)=\zeta(3) \sum_{\gamma\in \Gamma/\Gamma_{\infty}}[Im (\gamma S)]^{3/2},
\end{equation}
where  $\gamma$ indicates a transformation in $\Gamma=SL(2,Z)$ modded out by the subgroup defined by $\Gamma_{\infty}=\left(\begin{array}{cc}  \pm 1& n\\ 0 & \pm 1  \end{array}\right)$.

 The function $f_0$ can also 
be expressed as

\begin{equation}
\label{aj}
f_0(S, {\bar S})=\sum_{(m,n)\not=(0,0)}{\phi_2^{3/2} \over \vert m + n S\vert ^3}
\end{equation}
and  has the small $\phi_2$ expansion

\begin{equation}
\label{al}
f_0=2\zeta(3) \phi_2^{3/2} + {2 \pi\over 3} \phi_2^{-1/2} + 8 \pi \phi_2^{1/2} \sum_{m=0, n\geq 1}\left\vert{ m \over n  }\right\vert e^{2 i \pi m n \phi_1} K_1(2\pi \vert m n\vert \phi_2),
\end{equation}
where $K_1 $ is a Bessel function.
This form  for the exact $R^4$ corrections satisfies SL(2,Z) invariance, reproduces the correct perturbative expansion and the non-perturbative corrections are of the expected form.
The full four-point Lagrangian, including the analogous corrections for the other modes (antisymmetric fields and scalars) is given in Ref. \cite{kehagias}.
The metric and dilaton equations of motion now become

\begin{eqnarray}
\label{am} 
\alpha_1 K + { \alpha_2} f_0(S, {\bar S}) K^4 & = &0,\\
{ \alpha_2} f_0(S, {\bar S})_{,S} K^4 & = & 0,\\
{ \alpha_2} f_0(S, {\bar S})_{,{\bar S}} K^4 & = & 0,
\label{an}
\label{ana}
\end{eqnarray}
where $\alpha_1=\alpha$ and ${\alpha_2}=- {3\over 2^4}(D-3)(D-2)(D-1)$.
Hence, it is clear that it is now possible to satisfy all  equations provided  extrema of $f_0$ exist. This is the case as the fixed points, $S=e^{i \pi/6}$ and $S=1$ in the fundamental domain, are necessarily extrema of $f_0$. We have checked numerically that, in either case,  $f_0 > 0$  and, therefore, substituting in eq.~(\ref{am}), we find $K_s=(-{\alpha_1 \over f_0{\alpha_2}})^{1/3}>0$, corresponding to a de Sitter solution. However, this solution is unstable  since, defining ${\bar f}(K)
= \alpha_1 K + f_0{\alpha_2} K^4 $, then ${\bar f}'(K_s)=-3\alpha_1 < 0$. There remains, of course, the possibility that stable solutions become possible when higher order curvatures are included.

For type IIA theories, the tree level effective Lagrangian  for $R^4$ terms is the same as for type IIB theories, eq.~(\ref{aca}), but, at one loop, the (CP-even) contribution changes sign and there  is a  further (CP -odd) term

\begin{equation}
{\cal L}_{3}={1\over 3.2^6} \left(\zeta(3)+{\pi^2 \over 3} \phi_2^{-2}\right)\left(t_8 t_8 - {1\over 8} \epsilon_{10} \epsilon_{10} - {1\over 4} \epsilon_{10} t_8\right) R^4.
\end{equation}

As for type IIB theories, $R^4$ terms  receive no perturbative corrections beyond one loop, whose contribution  amounts to the exchange of eq.~(\ref{ah}); however, for type IIA theories there are  no non-perturbative corrections  in ten-dimensional type IIA theories  \cite{green2}. Hence, given the analysis presented above, it is not possible to obtain de Sitter (anti-de Sitter) solutions for these theories.

Finally, we turn to the question of whether the de Sitter solutions we found for type IIB theories up to order $\alpha'^3$ survive when higher oreder curvatures are taken into account. One expects that the  equations of motion (\ref{am})--(\ref{ana})  generalise to 

\begin{eqnarray}
\alpha_1 K + \alpha_2\ f_0(S, {\bar S})\ K^4 + \ldots + \alpha_{n-2}\ f_{n}(S, {\bar S})\ K^{n+4} + \ldots &=&0,\\
f_0(S, {\bar S})_{,S}\ \alpha_2\ K^4 + \ldots + \alpha_{n-2}\ f_n(S, {\bar S})_{,S}\ K^{n+4} + \ldots&=&0,\\
f_0(S, {\bar S})_{,\bar S}\ \alpha_2\ K^4 + \ldots + \alpha_{n-2}\ f_{n,\bar S}(S, {\bar S})_{,\bar S}\ K^{n+4} + \ldots&=&0,
\label{ba}
\end{eqnarray}

 Hence, it is clear  that it is possible to satisfy all  equations if the functions $f_n$  have at least one common extremum. Normally this would not happen but, for the case of type IIB superstrings, the SL(2,Z) symmetry requires  the functions $f_n$ to be  SL(2,Z)-invariant, implying that  they have common  fixed points, $S=e^{i \pi/6}$ and $S=1$, and these are then necessarily extrema of these functions \cite{shapere}. Hence, we conclude that maximally symmetric solutions exist, in principle,  for the superstring II  at all orders in $\alpha'$.
 Although this result is quite general and does not not depend on the particular form of the functions $f_n$  but solely on the assumption of SL(2,Z)-invariance, it is relevant  to mention  that the structure of  the higher order curvature terms has  already been discussed in the literature \cite{tseytlin1, partouche}, and a proposal exists for their form \cite{partouche}

\begin{equation}
\int d^{10} x \sqrt{-g} f_{-\frac{m}{2},0}(S,\bar S)\ R^{3m+1}
\end{equation}
where 

\begin{equation}
f_{s,k}(S, \bar S)=\sum_{(m,n)\not=(0,0)}{\phi_2^s\over (m S +n)^{s+k} (m\bar S + n)^{s-k}}
\end{equation}
are non-holomorphic modular forms of weights $(k, -k)$; hence, the functions $f_{-\frac{m}{2},0}$ are indeed modular invariant. Notice that $f_{3/2,0}$ is identical to $f_0$ introduced before. For $n\geq 1$ we have $s=3/2 -2 n<0$ and the infinite sum does not converge but can be defined by analytic continuation as

\begin{equation}
f_{s,k}(S,\bar S)=\pi^{2s-1} {\Gamma(1-s+k)\over \Gamma(s+k)} f_{1-s,k}(S,\bar S)
\end{equation}

 Hence, we see that the existence of de Sitter solutions for type IIB superstrings seems to be intimately related to the SL(2,Z) symmetry of the theory. In this context, it is interesting to notice  that, although stabilization of the dilaton and sucessfull inflation are very difficult features to implement for generic string cosmological models \cite{binetruy},  the situation seems to improve significantly for S-field  potentials that are  based on the assumption of SL(2,Z) invariance in the context of $N=1$ supergravity \cite{bento2}.

We conclude that inclusion of non-perturbative effects in type IIB superstrings
make it possible to obtain de Sitter solutions to the effective action at all orders in  $\alpha'$. This result cannot be extended to the type IIA theory, where the absence of non-perturbative corrections at this order precludes the existence of such solutions.

\vbs

We are grateful to F. Quevedo for important discussions on various aspects of this paper and, in particular, on the properties of modular-invariant functions. We also thank  A. Kehagias and O. Bertolami  for useful discussions on the subject of this paper. 

\end{document}